\documentstyle[11pt,newpasp,twoside,epsf]{article}
\footnotesize
\newdimen\digitwidth    
\setbox0=\hbox{\rm0}
\digitwidth=\wd0
\catcode`!=\active
\def!{\kern\digitwidth}
\normalsize

\begin{document}

\title{Timing the Parkes Multibeam Pulsars}

\author{R. N. Manchester$^1$, A. G. Lyne$^2$, F. Camilo$^2$, 
V. M. Kaspi$^3$, I. H. Stairs$^2$, F. Crawford$^3$, D. J. Morris$^2$,
J. F. Bell$^1$, N. D'Amico$^4$}
\affil{$^1$Australia Telescope National Facility, CSIRO, PO Box 76, Epping NSW
1710, Australia}
\affil{$^2$University of Manchester, Jodrell Bank Observatory, Macclesfield,
Cheshire SK11~9DL, UK}
\affil{$^3$Center for Space Research, MIT, Cambridge MA 02139, USA}
\affil{$^4$Osservatorio Astronomico di Bologna, 40127 Bologna, Italy}

\begin{abstract}
Measurement of accurate positions, pulse periods and period derivatives is
an essential follow-up to any pulsar survey. The procedures being used to
obtain timing parameters for the pulsars discovered in the Parkes multibeam
pulsar survey are described. Completed solutions have been obtained so far
for about 80 pulsars. They show that the survey is preferentially finding
pulsars with higher than average surface dipole magnetic fields. Eight
pulsars have been shown to be members of binary systems and some of the more
interesting results relating to these are presented.
\end{abstract}

\section{Introduction}
The Parkes multibeam pulsar survey is proving to be extraordinarily
successful, with more than 400 pulsars discovered so far (Lyne et al. 1999;
Camilo et al., this volume, astro-ph/9911185). These pulsars provide a magnificent database for many
different studies related to pulsar formation and evolution, the pulse
emission process and the interstellar medium. The feasibility of essentially
all of these studies is dependent on knowledge of accurate positions, pulse
periods and period derivatives. These parameters are normally obtained from pulse
timing observations made over a period of 12 months or more. Such
observations also reveal the presence of various perturbations to the
period, notably the effect of orbital motion. Characterisation of such
orbital motion is an important part of any timing program.

Obtaining these timing parameters for the Parkes multibeam pulsars is a
major task which we share between the Parkes and Jodrell Bank observatories.
We first briefly describe our observational procedures and then discuss some
trends in the distribution of period derivatives from the results obtained
so far. Finally we highlight some of the interesting features of the eight
binary pulsars so far detected.

\section{Observing and Analysis Procedures}
Observations to measure the basic timing parameters of the pulsars
discovered in the Parkes multibeam survey are being made using both the
Parkes 64-m radio telescope of the Australia Telescope National Facility,
and the 76-m Lovell Telescope of Jodrell Bank Observatory. With a few
exceptions, pulsars north of declination $-35\deg$ are being timed at
Jodrell Bank, whereas those south of this declination are being timed at
Parkes. At Parkes, the centre beam of the multibeam system is used for the
timing obserations. As for the survey, a 288-MHz bandwidth centred on 1374
MHz is observed in two orthogonal polarizations; the system temperature for
the centre beam at high Galactic latitudes is about 21 K. At Jodrell Bank, a
96-MHz bandwidth centred on 1376 MHz is observed with a dual-polarization
receiver having a system temperature of about 30 K at high Galactic
latitudes. Filterbank channel bandwidths are 3 MHz at both observatories. 

At Parkes, the data are one-bit digitized and written to Exabyte tape for
subsequent processing. This consists of synchronously folding the data at
the topocentric pulsar period to form sub-integrations of typical duration 1
-- 2 minutes. These are then dedispersed and stored as archive files on
disk. The same procedure is followed at Jodrell Bank, except that the
folding and dedispersing are done on-line. Observing times are adjusted to
give signal-to-noise ratios for final average profiles which are typically
between 10 and 20. For each pulsar, profiles obtained in this way are
cross-correlated with a standard template to give pulse times-of-arrival
(TOAs). These are then analysed using the timing programs TEMPO (see
http://pulsar.princeton.edu/tempo) or, at Jodrell Bank, PSRTIME.

Observations are made over a period of 12 -- 18 months following
confirmation at intervals typically of 4 -- 8 weeks, with some more closely
spaced observations to resolve pulse-counting ambiguities. These
observations also reveal pulsars with unusual timing behaviours, in
particular, those which are members of binary systems. These are observed
more intensively to determine their characteristics.

Once a satisfactory timing solution is obtained, the pulsar is given its
official Jname and the parameters are entered in the pulsar catalogue. At
this time, the parameters are also made available on the Parkes multibeam
pulsar survey web pages (see Bell et al., this volume, astro-ph/9911321). Timing
observations for these pulsars are then discontinued except in the case of
binary or other pulsars of especial interest.

\section{Slow-down Rates and Implications}
Full timing solutions have been obtained for about 80 of the pulsars
discovered so far. Fig.~1 is a plot of period derivative versus pulse period
showing the multibeam pulsars, previously known radio pulsars in the
Galactic disk (i.e., excluding pulsars associated with globular clusters)
and anomalous X-ray pulsars (AXPs) with known period derivatives
(e.g. Mereghetti, Israel \& Stella 1998). Multibeam pulsars are much more
concentrated toward the top of the region occupied by `normal'
(non-millisecond) pulsars on the $P - \dot P$ plane. Since, on average, the
multibeam pulsars are much more distant than previously known pulsars
(Camilo et al., this volume), this bias indicates a correlation between
spin-down rate and radio luminosity.

It is notable that the three radio pulsars with the highest known surface
dipole magnetic fields have been discovered in this survey. These pulsars
have surface fields in the range $(2.3 - 5.5) \times 10^{13}$ G, beyond the
point where some models predict radio emission should cease (e.g. Baring \&
Harding 1998). In particular, the multibeam pulsar PSR J1814$-$1744 (pulse
period 3.97 s) lies very close to the AXP 1E 2259+586 on the $P - \dot P$
plane. No radio emission has been detected from this AXP (Coe, Jones \&
Lehto 1994). This suggests that the underlying reasons why a neutron star
manifests itself as a radio pulsar or as an AXP are not simply dependent on
the pulsar spin period and implied surface dipole magnetic field.

\begin{figure}[ht]
\plotfiddle{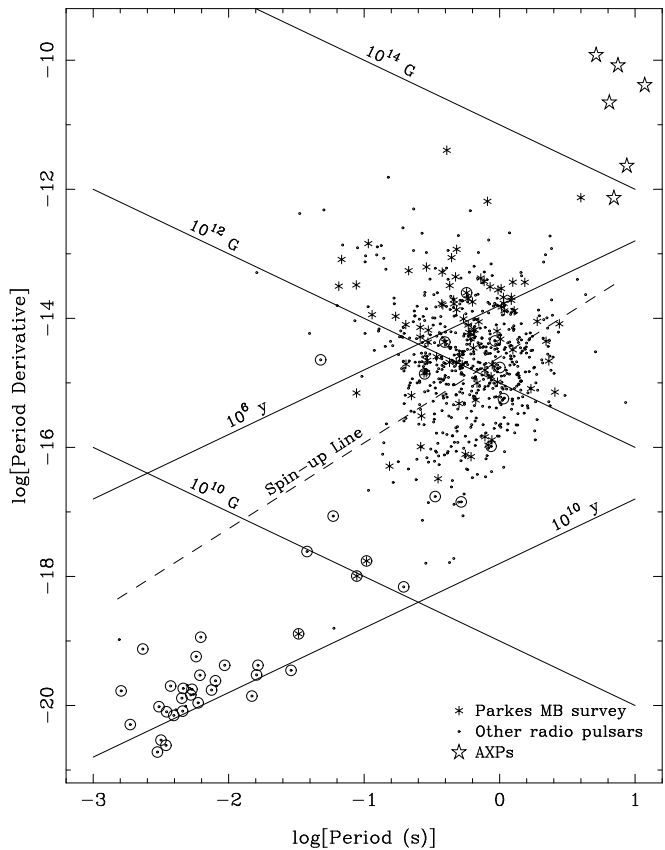}{105mm}{0}{52}{52}{-170}{-70}
\caption{Plot of period derivative versus pulse period for the Parkes
multibeam survey pulsars that have full timing solutions and for previously
known pulsars in the Galactic disk. Pulsars that are members of a binary
system are marked with a circle. The large open stars represent anomalous
X-ray pulsars. Lines of constant surface dipole magnetic field and pulsar
characteristic age are marked. The dashed line is the minimum period for
spin-up by accretion from a binary companion (Bhattacharya \& van den Heuvel
1991).}
\end{figure}

Another of these high-$B$ pulsars, PSR J1119$-$6127, has the much shorter
period of 0.407 s and hence is very young; its characteristic age, $\tau_c =
P / (2 \dot P)$, is about 1600 years. We have already measured what appears
to be a significant braking index for this pulsar, $3.0 \pm 0.1$. This is
only the fourth measured braking index and the only one consistent with the
magnetic-dipole value of 3.0. Future observations will show whether or not
this value is stable and truly representative of the secular slow-down.

\section{Binary Pulsars}

So far, eight of the multibeam pulsars have been shown to be members of
binary systems. The basic parameters for these pulsars are shown in Table
1. Distances are estimated from the dispersion measures using the Taylor \&
Cordes (1993) electron density model and minimum companion masses are
derived from the mass function assuming a pulsar mass of 1.4 $M_{\sun}$. It
is notable that all of these systems have relatively high-mass
companions. They naturally split into three categories: low-eccentricity
systems, high-eccentricity systems of intermediate mass, and high-mass
systems. None of the low-eccentricity systems has a companion mass of less
than 0.15 $M_{\sun}$ whereas about half of the previously known similar
systems have minimum companion masses less than this value. Two of the
multibeam pulsars have the largest known minimum companion masses for
low-eccentricity systems.

\begin{table}[ht]
\caption{Parameters for Parkes multibeam binary pulsars}
\begin{tabular}{lcccccc} \hline
PSR J & $P$ & $\tau_c$ & Distance & $P_b$ & Ecc. & Min. $M_c$ \\
      & (ms)& ($10^6$ y) & (kpc) & (d) &  & ($M_\odot$) \\ \hline
1232$-$6501 &   !88.28 & 1400 &   10.0 &    !!1.863 &   0.00 &    !0.15 \\
1904+04     &   !71.09 & --  &   !4.0 &    !15.750 &   0.04 &    !0.23 \\ 
1810$-$2005 &   !32.82 & 4000  &   !4.0 &    !15.012 &   0.00 &    !0.29 \\
1453$-$58   &   !45.25 & --  &   !3.3 &    !12.422 &   0.00 &    !0.88 \\
1435$-$60   &   !!9.35 & --  &   !3.2 &    !!1.355 &   0.00 &    !0.90 \\ \\
1811$-$1736 &   104.18 & !950  &   !5.9 &    !18.779 &   0.83 &    !0.87 \\
1141$-$6545 &   393.90 & !!!1.45  &   !3.2 &    !!0.198 &   0.17 &    !1.01 \\ \\
1740$-$3052 &   570.31 & !!!0.36  &   10.8 &    231.039 &   0.58 &    11.07 \\ \hline
\end{tabular}
\end{table} 	 

The next category of binary system, with high eccentricity and companion
mass about 1 $M_{\sun}$, has two members in the multibeam sample. The first,
PSR J1811$-$1736, has all the hallmarks of a double neutron-star system
(Lyne et al. 1999). It has a very large characteristic age, suggesting a
history of recycling, a moderately long pulse period, high eccentricity and
parameters consistent with a 1.4 $M_{\sun}$ companion. The other system, PSR
J1141$-$6545, only recently discovered, is superficially similar, but has
some important differences. Its orbital period is very short, only 4.8 h. An
accurate position has been obtained for the pulsar using the Australia
Telescope Compact Array and this has allowed a measurement of the period
derivative from the $\sim 5$-week data span presently
available. Significantly, the characteristic age is relatively short at
$\sim 1.5$ My. We have also already obtained a significant measurement of
the orbital precession, $5.5 \pm 0.4$ \deg/y, the highest known value. If
general relativistic precession is the only significant effect, this implies
a total mass for the system of $2.4 \pm 0.2 M_{\sun}$. These are certainly
preliminary results, but they suggest that the companion may be a heavy
white dwarf, similar to the PSR B2303+46 system (van Kerkwijk \& Kulkarni
1999), rather than a neutron star.

The final system in the list, PSR J1740$-$3052, is distinguished by its very
massive companion, with a minimum mass $\sim 11 M_{\sun}$. As Fig. 2 shows, the
orbit is highly eccentric. The very large mass restricts the possible types
of companion star to a massive `normal' (i.e. non-degenerate) star or a
black hole. We are currently attempting to determine which of these two
possiblities is correct. Unfortunately, the pulsar lies at low Galactic
latitude within $2.2\deg$ of the Galactic Centre, and is probably at least as
distant. Optical searches are therefore futile, even for an O-B star. We
have obtained infrared K-band images of the region using the 2.3-m telescope
at Siding Spring Observatory and the 3.9-m Anglo-Australian Telescope which
show that the pulsar is coincident within 0\farcs4 with a star having a
K-band magnitude of about 11. Infrared spectra of the star are consistent
with its being a K5 supergiant, which could be sufficiently massive to be
the pulsar companion. They also show Brackett-$\gamma$ in emission, which
suggests irradiation of the star, for example, by a pulsar companion.

\begin{figure}[t]
\plotfiddle{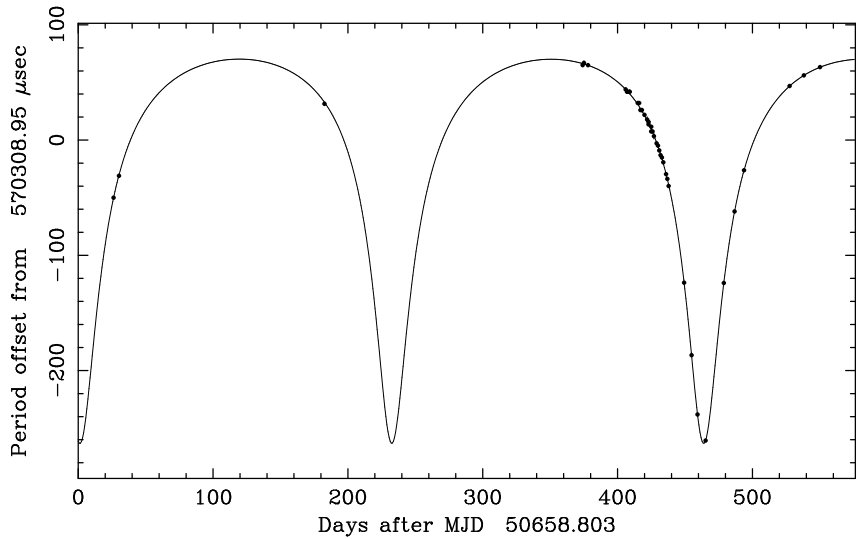}{65mm}{270}{45}{45}{-185}{230}
\caption{Period variations for PSR J1740$-$3052 over a 500-day
interval. Observed points are marked by dots; the period uncertainty is less
than the size of the dot. The smooth curve represents a fit to the data of
an orbit with period of 231 days. }
\end{figure}

However, if this star is the companion, it is surprising that there is no
evidence for any eclipse of the pulsar emission -- the pulsar passes within
1.25 nominal stellar radii at periastron. Fig. 2 shows that the
pulsar was detectable at several times very close to periastron
passage. We plan further observations to search for the expected effects of
a close encounter of the pulsar with a supergiant companion.

\section{Conclusions}
The Parkes multibeam pulsar survey is living up to its promise. It is
finding large numbers of pulsars, many of which are young and hence
especially interesting for studies of neutron star evolution and
related topics. It is also finding some very interesting binary
pulsars, the study of which will have a significant impact on our
understanding of binary and stellar evolution. The survey is currently
only about 50 per cent complete and timing solutions have been
obtained for only about 15 per cent of the expected final complement
of pulsars. We certainly hope and expect that many more interesting and
important objects will be uncovered by this survey.

\acknowledgements Many of our colleagues have contributed to the
observations described here. We particularly thank the staff of the Parkes
Observatory for their support, and Stuart Lumsden, Lowell Tacconi-Garman,
Peter Wood, Russell Cannon and Nigel Hambly for assistance with the optical
and infrared observations and analysis of the PSR J1740$-$3052 system. The
Parkes radio telescope is part of the Australia Telescope which is funded by
the Commonwealth of Australia for operation as a National Facility managed
by CSIRO.

\end{document}